\documentclass[12pt]{article}
\usepackage{amsmath,amssymb}
\usepackage{graphicx,color}
\numberwithin{equation}{section}
\usepackage{cite}
\usepackage{bm}
\usepackage{dcolumn}
\newcommand{\bea}{\begin{eqnarray}}
\newcommand{\eea}{\end{eqnarray}}
\newcommand{\nn}{\nonumber}
\newcommand{\na}{\nabla}

\begin{document}
\author{\large Taeyoon
Moon\footnote{dpproject@skku.edu}\addtocounter{footnote}{2}\,,
Phillial Oh\footnote{ploh@skku.edu}\addtocounter{footnote}{-2}, and
Jongsu Sohn\footnote{jongsusohn@skku.edu}\\
{\it  \small Department of Physics and
Institute of Basic Science, }\\
{\it \small Sungkyunkwan University, Suwon 440-746 Korea}}
\title{Anisotropic Weyl symmetry and cosmology}
\maketitle

\begin{abstract}
We construct an anisotropic Weyl invariant theory in the ADM
formalism and discuss its cosmological consequences. It extends the
original anisotropic Weyl invariance of Ho\v{r}ava-Lifshitz gravity
using an extra scalar field. The action is invariant under the
anisotropic transformations of the space and time metric components
with an arbitrary value of the critical exponent $z$. One of the
interesting features is that the cosmological constant term
maintains the anisotropic symmetry for $z=-3$. We also include the
cosmological fluid and show that it can preserve the anisotropic
Weyl invariance if the equation of state satisfies $P= z\rho/3$.
Then, we study cosmology of the Einstein-Hilbert-anisotropic Weyl
(EHaW) action including the cosmological fluid, both with or without
anisotropic Weyl invariance. The correlation of the critical
exponent $z$ and the equation of state parameter $\bar{\omega}$
provides a new perspective of the cosmology. It is also shown that
the EHaW action admits a late time accelerating universe for an
arbitrary value of $z$ when the anisotropic conformal invariance is
broken, and the anisotropic conformal scalar field is interpreted as
a possible source of dark energy.
\end{abstract}

\section{Introduction}

Gravity theory with a local Weyl invariance was proposed as an
alternative theory of gravity \cite{weyl} and various aspects have
been investigated for a long time. Among them, there are two main
avenues. The first one is conformal gravity where the theory is
built on the local conformal invariance and the general covariance.
In this theory, the conformal invariant action can be realized by
introducing the quadratic Weyl curvature tensor. This theory has the
dimensionless gravitational coupling constant and the property that
it is renormalizable \cite{stelle}, asymptotically free
\cite{fradkin} and could be potentially unitary \cite{bender}. The
other way of achieving conformal invariance is to introduce an extra
gauge scalar field to compensate the non-invariance of the
Einstein-Hilbert action \cite{dirac}. Ever since,  many attempts of
incorporating conformal invariance in the theory of general
relativity have been carried out in the diverse areas of theoretical
physics ~\cite{cheng,hehl,pawl,nishino1,kell,demir,jackiw,jain1}.

 One application of conformal symmetry in the latter case is to consider the conformal
  scalar field which is non-minimally coupled  to the curvature scalar with a
coupling constant \cite{birrell}. In such a theory, the gravity
sector is described by the Einstein-Hilbert action, and the
conformal scalar field is treated as conformal matter. The exact
conformal symmetry is imposed with a specific choice of the
non-minimal coupling constant. In particular, this approach found
wide applications in cosmological models
\cite{birrell,qu1,wetterich} in which it was first proposed to
describe decaying cosmological constant \cite{dolgov}.


The original conformal invariance is isotropic in the sense that the
time and space components of the metric transform in the same
manner. On the other hand, the canonical ADM formalism
\cite{Arnowitt} decomposes space-time into space and time. In this
background, one could envisage  an anisotropic Weyl transformation
in which the space and time components of the metric transform
differently, but still leaves the ``action'' invariant. Especially,
such an attempt is well motivated by the recent upsurge \cite{hl} of
interest in the Ho\v{r}ava-Lifshitz gravity \cite{Horava} which has
the feature of anisotropy between the space and time. In fact, in
\cite{Horava} it is shown that classical action can have an
anisotropic (local) Weyl invariance for specific values of the
critical exponent $z=3$ and the free parameter of the metric on the
space of metrics. In this case, each component of the metric
transforms as $ g_{00}\rightarrow e^{6\omega(t,x)}g_{00}~,
 g_{0i}\rightarrow e^{2\omega(t,x)}g_{0i}~, g_{ij}\rightarrow e^{2\omega(t,x)}g_{ij}
 .$

As was mentioned before, however, one can also construct a
conformally invariant gravity with curvature scalar by introducing
an extra scalar field. Therefore, it seems natural to attempt to
extend the anisotropic Weyl invariance of Ho\v{r}ava-Lifshitz
gravity in this way. The purpose of this work is to show that this
type of extension is indeed possible. We first review the
anisotropic Weyl invariance of Ho\v{r}ava-Lifshitz gravity in the
ADM formalism. Then, we extend the analysis to anisotropic Weyl
invariant theory by introducing a scalar field with a suitable
conformal weight which is given by the critical exponent.
 It turns out that the conformal invariance can be preserved
for any value of the critical exponent, especially under the
transformation $g_{00}\rightarrow e^{2z\omega(t,x)}g_{00}$.
We will treat the resulting anisotropic conformal scalar field as
describing the conformal matter and couple to  the Einstein-Hilbert
action\footnote{ In this paper, we only pay attention to the
anisotropic Weyl invariance at the lowest curvature level and do not
include the higher derivative terms such as the Cotton tensor, $R^2$
term, etc. in the action.}. For $z=1$, the action reduces to the
conformal matter theory of Ref. \cite{birrell,qu1,wetterich}.
Otherwise, the local Lorentz invariance is explicitly broken.

One of the motivations for considering an arbitrary value of $z$ is
that we look for the role of the critical exponent in the cosmology
and interpret  the anisotropic conformal scalar field as a possible
source of dark energy \cite{copeland}. In order to describe the
cosmology in this context, we first break the conformal invariance
\cite{vet,jain,nishino} explicitly by considering an arbitrary
potential term for the scalar field. Then, we search for
cosmological solutions paying attention to the possible role of $z$
in the evolution of the universe. For example, we compare with the
cosmological model with $z=1$, and show that accelerating phase can
exist in vacuum with an arbitrary value of $z$.  We also introduce a
cosmological fluid with equation of state $\bar{\omega}$, and study
the cosmology. If the conformal invariance is imposed on the fluid,
we find that the critical exponent and the equation of state must be
correlated. When the conformal invariance is broken, there is no
correlation and the critical exponent $z$ remains as a free
parameter.  We will present
 the conformal preserving case also, because it seems that the
cosmology in this case also shows some interesting feature. For
example, the cosmological constant term enjoys anisotropic Weyl
invariance with $z=-3$, in which case a cosmological solution which
extrapolates between the matter dominated epoch and the accelerating
phase exists.

This paper is organized as follows. In Sec.2, we briefly review the
anisotropic Weyl invariance in the Ho\v{r}ava-Lifshitz gravity. In
Sec.3, we consider the Einstein-Hilbert action with an (isotropic)
conformal symmetry in the ADM formalism and  extend to the action
with an anisotropic Weyl invariance by introducing a scalar field.
In particular, we show that cosmological fluid can preserve the
anisotropic Weyl invariance if the equation of state satisfies $P=
z\rho/3$. In Sec.4, we study cosmology with the FRW metric and apply
it to the EHaW action with an arbitrary potential including also the
cosmological fluid with or without anisotropic Weyl invariance. We
briefly summarize the results and discuss them in Sec.5.

\section{Anisotropic Weyl invariance in the Ho\v{r}ava-Lifshitz gravity}

Let us consider an action of $z=3$ gravity theory in $1+3$
dimensions \cite{Horava}: \bea S_{aH}&=&\int
dt\,d^3x\,\sqrt{g}\,N\left\{\frac{2}{\kappa^2}\left(
K_{ij}K^{ij}-\lambda K^2\right)-\frac{\kappa^2}{2w^4}C_{ij}C^{ij}
\right\},\label{HL}\eea where $\kappa,~w$ are dimensionless constant
parameters, $K_{ij}$ is the extrinsic curvature which is defined by
\begin{eqnarray}
K_{ij} = -\frac{1}{2N} (\partial_t g_{ij} - \nabla_i N_j - \nabla_j
N_i)\,
\end{eqnarray}
and
\begin{eqnarray}
C^{ij}=\epsilon^{ik\ell}\nabla_k\left(R^j{}_\ell
-\frac14R\delta_\ell^j\right)
\end{eqnarray}
is the Cotton tensor. Here, $R$ is the curvature scalar in 3 space
dimensions. Under anisotropic Weyl transformation,
\begin{eqnarray}
 N\rightarrow
e^{3\omega(t,x)}N,~~ N_i\rightarrow e^{2\omega(t,x)}N_i,~~
 g_{ij}\rightarrow e^{2\omega(t,x)}g_{ij},\label{KK}
\end{eqnarray}
the action (\ref{HL}) transforms to
\begin{eqnarray}
S_{aH}&\rightarrow& S_{aH}=\int
dt\,d^3x\,\sqrt{g}\,N\left\{\frac{2}{\kappa^2}\left[
K_{ij}K^{ij}-\lambda
K^2-2(1-3\lambda)K\left(\frac{\dot{\omega}-\nabla_i\omega
N^{i}}{N}\right)\right.\right.\nn\\
&&~~~~~~~~~~~~~~~~~~~~+\left.\left.3(1-3\lambda)\left(\frac{\dot{\omega}-\nabla_i\omega
N^{i}}{N}\right)^2\right]-\frac{\kappa^2}{2w^4}C_{ij}C^{ij} \right\}.\label{qqq}
\end{eqnarray}
In the above equation (\ref{qqq}), the third and fourth terms vanish
with $\lambda=1/3$, and the action (\ref{HL}) has an anisotropic
Weyl invariance. As was shown in \cite{Horava}, this symmetry is an
additional gauge symmetry supplementing the foliated
diffeomorphisms.

 It is interesting to note that under general
anisotropic Weyl transformation,
\begin{eqnarray}
 N\rightarrow
e^{z\omega(t,x)}N,~~ N_i\rightarrow e^{2\omega(t,x)}N_i,~~
 g_{ij}\rightarrow e^{2\omega(t,x)}g_{ij},\label{KK1}
\end{eqnarray}
the extrinsic curvature terms become
\begin{eqnarray}
&&K_{ij}K^{ij}-\frac{1}{3} K^2\rightarrow
e^{-2z\omega}(K_{ij}K^{ij}-\frac{1}{3} K^2).\label{KT}
\end{eqnarray}
Since the volume element transforms as $\sqrt{g}N\rightarrow
e^{(3+z)\omega}\sqrt{g}N$ under the transformation $(\ref{KK1})$, it
has no anisotropic Weyl invariance unless $z=3$.  One may, however,
write an anisotropic Weyl invariant action by introducing some
scalar field which can compensate the conformal weight $3-z$. With a
proper power of this scalar field, the conformal weight $z-3$ coming
from the volume element and the Cotton tensor term which transforms
as $C_{ij}C^{ij}\rightarrow e^{-6\omega}C_{ij}C^{ij}$, can also be
compensated. In the next section, we will explicitly construct
anisotropic Weyl action which is invariant for an arbitrary $z$
including curvature scalar. As was mentioned before, we will drop
the Cotton tensor term and include only terms up to second
derivatives.

\section{Anisotropic Weyl invariant gravity for general $z$}

Let us first consider a conformally invariant action in four
dimensions. This is given by
\begin{eqnarray}
S_C=\int \,
d^4x\sqrt{-g}\phi^2\left(R_{(4)}-6\frac{\na_{\gamma}\na^{\gamma}\phi}{\phi}\right).
\label{sc}
\end{eqnarray}
The above action is invariant under \cite{dirac,kell,con}
\begin{eqnarray}
g_{\mu\nu}\rightarrow e^{2\omega}g_{\mu\nu},~~\phi\rightarrow
e^{-\omega}\phi,
\end{eqnarray}
 where $\omega=\omega(t,x)$.
Considering the ADM decomposition and rearranging terms, it can be
rewritten as~\cite{kell}
\begin{eqnarray}
S_C=\int \, dt d^3x N\sqrt{g}~\varphi^{4}\left(R-8
\frac{\nabla_{i}\nabla^{i}\varphi}{\varphi}+B_{ij}B^{ij}-B^2\right),
\label{conformalADM}
\end{eqnarray}
where $\varphi^2=\phi$ and
\begin{eqnarray}
B_{ij}&=&K_{ij}-\frac{2}{N\varphi}g_{ij}(\dot{\varphi}-\na_{i}\varphi
N^{i})\\&\equiv&K_{ij}+\frac{\theta}{2N} g_{ij},
\end{eqnarray} with
$\theta=-4(\dot{\varphi}-\nabla_{i}\varphi N^{i})/\varphi$. The
above action (\ref{conformalADM}) is invariant under
\begin{eqnarray}
 N&\rightarrow&
e^{\omega}N,~~~N_{i}\rightarrow e^{2\omega}N_{i},\nn\\
g_{ij}&\rightarrow& e^{2\omega}g_{ij},~~\varphi\rightarrow
e^{-\frac{\omega}{2}}\varphi.
\end{eqnarray}

We can extend the above procedure to  an anisotropic Weyl
invariance. One can show that the following action
\begin{eqnarray}
S_{\varphi}=\int \,dt
d^3xN\sqrt{g}~\left[~\varphi^{2z+2}\left(R-8\frac{\nabla_{i}\nabla^{i}\varphi}{\varphi}\right)+
\varphi^{-2z+6}B_{ij}B^{ij}
-\varphi^{-2z+6}B^2~\right],\label{conformalR}
\end{eqnarray}
 is invariant with respect to \bea N&\rightarrow&
e^{z\omega}N,~~~N_{i}\rightarrow e^{2\omega}N_{i},\nn\\
g_{ij}&\rightarrow& e^{2\omega}g_{ij},~~\varphi\rightarrow
e^{-\frac{\omega}{2}}\varphi. \label{conformalq}\eea Note that we
have a factor $z$ in the above equations which turns out to be
coincident with the critical exponent. That is, beside the local
transformations (\ref{conformalq}), the above action
(\ref{conformalR}) is also invariant with respect to a global
transformation \bea t\rightarrow b^z t,~~x\rightarrow
bx,~~N_{i}\rightarrow b^{1-z}N_{i},~~ \varphi\rightarrow
b^{-\frac{1}{2}}\varphi,\eea with $N$ and $g_{ij}$ being unchanged.
When $z=1$, the above action (\ref{conformalR}) reduces to the ADM
decomposition of the conformally invariant action (\ref{sc}). On the
other hand, in the same manner, considering the transformation law
(\ref{KT}) in the previous section, the extrinsic curvature terms
with an anisotropic Weyl invariance can be written as
\begin{eqnarray}
S_{aK}&=&\int dt\,d^3x\,\sqrt{g}\,N\left\{\varphi^{-2z+6}\left(
K_{ij}K^{ij}-\frac{1}{3} K^2\right)\right\}.\label{CK}
\end{eqnarray}

From the action (\ref{conformalR}), (\ref{CK}), one can construct the general
action with anisotropic Weyl invariance as follows
\begin{eqnarray}
S_{aW}&=&\int \,dt
d^3xN\sqrt{g}~\left[~\varphi^{2z+2}\eta\left(R-8\frac{\nabla_{i}\nabla^{i}\varphi}
{\varphi}\right)+
\right.\nn\\
&&\left.\varphi^{-2z+6}(\eta+\xi)B_{ij}B^{ij}
-\varphi^{-2z+6}(\eta+\frac{\xi}{3})B^2-V(\varphi)\right],\label{conformal}
\end{eqnarray}
where $\eta,~ \xi$ are some constants and for the anisotropic Weyl
invariance, $V(\varphi)=\alpha\varphi^{2(z+3)}$ with some constant
$\alpha$. Note that in the case of $z=1$ the transformation law of
$N$ becomes isotropic Weyl transformation, i.e., ordinary conformal
transformation. As expected, for $\eta=1$, $\xi=\alpha=0$ and
$\eta=\alpha=0$, $\xi=1$ the action (\ref{conformal}) becomes
(\ref{conformalR}) and (\ref{CK}) respectively. We also note that in
the case of $z=-3$, the potential term becomes a cosmological
constant.

The most general action which includes the cosmological fluid will
be given by
\begin{eqnarray}
S_{aWf}=S_{aW}+S_f,\label{cconformal}
\end{eqnarray}
where $S_f$ is the action which has anisotropic Weyl invariance
without $\varphi$ term. We show in the following lines that the
equation of the state for this cosmological fluid satisfies
\begin{eqnarray}
P=\frac{z}{3}\rho \label{zw}.
\end{eqnarray}
In order to see this, we first vary for $N$, $N^{i}$, $g^{ij}$,
$\varphi$, and obtain the equations of motion:
\begin{eqnarray}
\delta_{N}S_{aWf};&&\varphi^{2z+2}\{\eta\left(R-8\frac{\nabla_{i}\nabla^{i}\varphi}{\varphi}\right)
            -\varphi^{-4z+4}(\eta+\xi)B_{ij}B^{ij}
+\varphi^{-4z+4}(\eta+\frac{\xi}{3})B^2\}\nn\\
&&-\alpha\varphi^{2(z+3)}-\rho=0,~~~\label{Neq}\\
\delta_{N^{i}}S_{aWf};&&2\varphi^{-2z+6}
                 \{-(\eta+\xi)\nabla_{j}B^{j}_{i}+2(z-3)(\eta+\xi)\frac{\nabla_{j}\varphi}{\varphi}
B^{j}_{i}\}\nn\\
&&+2\varphi^{-2z+6}\{(2(1-z)\eta+\frac{2}{3}(3-z)\xi)\frac{\nabla_{i}\varphi}{\varphi}
B+(\eta+\frac{\xi}{3})\nabla_{i}B\}=0,\label{Nieq}\\
\delta_{g^{ij}}S_{aWf};&&N\varphi^{2z+2}\{\eta
A_{ij}^{(1)}+\varphi^{-4z+4}(\eta+\xi)A_{ij}^{(2)}-\varphi^{-4z+4}(\eta+\frac{\xi}{3})A_{ij}^{(3)}\}\nn\\
&&+\frac{N}{2}
\alpha\varphi^{2(z+3)}g_{ij}-\frac{N}{2}g_{ij}P=0,\label{gijeq}
\end{eqnarray}
where $\rho=-\left.\frac{1}{\sqrt{g}}\frac{\delta S_f}{\delta
N}\right., ~Pg_{ij}=-\left.\frac{2}{N\sqrt{g}}\frac{\delta
S_f}{\delta g^{ij}}\right.$,
\begin{eqnarray}
A_{ij}^{(1)}&=&R_{ij}-\frac{1}{2}g_{ij}R-\frac{\nabla_{i}\nabla_{j}N}{N}
-4(z-1)\frac{\nabla_{i}N}{N}\frac{\nabla_{j}\varphi}{\varphi}+4z\frac{\nabla_{j}N}{N}\frac{\nabla^{j}
\varphi}{\varphi}g_{ij}\nn\\
&&-2(2z+1)(z-3)\frac{\nabla_{i}\varphi\nabla_{j}\varphi}{\varphi^2}+\frac{\nabla_{k}\nabla^{k}N}{N}g_{ij}+
2(z-1)(2z+1)\frac{\nabla_{k}\varphi\nabla^{k}\varphi}{\varphi^2}g_{ij}-\nn\\
&&2(z+1)\frac{\nabla_{i}\nabla_{j}\varphi}{\varphi}+
2(z+1)\frac{\nabla_{k}\nabla^{k}\varphi}{\varphi}g_{ij},\nonumber\\
A_{ij}^{(2)}&=&\frac{-N_i\nabla_{k}B_{j}^{k}}{N}-\frac{N_j\nabla_{k}B_{i}^{k}}{N}+\frac{\nabla_{i}N^{k}B_{jk}}{N}
+\frac{\nabla_jN^{k}B_{ik}}{N}+\frac{N_{k}\nabla^{k}B_{ij}}{N}-2B_{i}^{k}B_{jk}-\frac{1}{2}B_{kl}B^{kl}g_{ij}+\nn\\
&&BB_{ij}-\frac{\dot{B_{ij}}}{N}+(1-\frac{z}{2})\frac{\theta
B_{ij}}{N}+4(z-3)\frac{\nabla_{k}\varphi}{\varphi}
\frac{N_{i}B^{k}_{j}}{N}+\frac{4\nabla_{i}\varphi}{\varphi}\frac{N_{j}B}{N}
,\nonumber\\
A_{ij}^{(3)}&=&\frac{B^2}{2}g_{ij}-\frac{\nabla_{j}BN_{i}}{N}
-\frac{\nabla_{i}BN_{j}}{N}+2z\frac{\nabla_{j}\varphi}{\varphi}
\frac{N_iB}{N}+2z\frac{\nabla_{i}\varphi}{\varphi}\frac{N_{j}B}{N}-\frac{z\theta}{2N}Bg_{ij}+
\frac{\nabla_{k}BN^{k}g_{ij}}{N}-\frac{\dot{B}g_{ij}}{N},\nonumber
\end{eqnarray}
and
\begin{eqnarray}
\delta_{\varphi}S_{aWf};&&N\varphi^{2z+1}[(2z+2)\eta
R-16(2z+1)\eta\frac{\nabla_{i}\nabla^{i}\varphi}{\varphi}-16z(2z+1)\eta\frac{\nabla_i\varphi
\nabla^{i}\varphi}{\varphi^2}
+\nn\\
&&\{2(\eta-\xi)+2z(\eta+\frac{\xi}{3})\}B^2\varphi^{-4z+4}-4z\eta
NB\theta\varphi^{-4z+4}+
8\eta\nabla_{i}BN^{i}\varphi^{-4z+4}+\nn\\
&&(-2z+6)(\eta+\xi)B_{ij}B^{ij}\varphi^{-4z+4}-8\eta\nabla_{i}\nabla^{i}N
-16(2z+1)\eta\nabla_{i}N\frac{\nabla^{i}\varphi}{\varphi}-8\eta\dot{B}\varphi^{-4z+4}]\nn\\
&&-2(z+3)N\alpha\varphi^{2z+5}=0. \label{phieq}
\end{eqnarray}
In the Appendix, we show that Eq. (\ref{zw}) must be satisfied in
order to be consistent with Eqs. (\ref{Neq})$\sim$(\ref{phieq}).
Note that this condition is only for cosmological fluid with
anisotropic Weyl invariance\footnote{When $z=1$, this condition
corresponds to $T^{\mu}_{\mu}=0$ which represents the condition of
isotropic conformally invariant fluid. That is,
 the above condition (\ref{zw}) is
 equivalent to $P=\frac{1}{3}\rho$.
  It is pointed out that the
 above condition (\ref{zw}) also can be obtained by dimensional
 analysis as shown in \cite{bin}.}. It can be violated, if we do not insist on
 the symmetry. In
the next section, we will consider cosmological consequences of both
unbroken and broken cases. In the broken case, we  consider an
arbitrary potential $V(\varphi)$ breaking anisotropic Weyl
invariance in Eq. (\ref{conformal}).

\section{Cosmological solutions}

In order to investigate cosmological consequences, we first recall
that the Einstein-Hilbert action in the ADM formalism is given by
\begin{eqnarray}
S_{EH}=\int \,dt
d^3xN\sqrt{g}~\frac{1}{2\kappa^2}(R+K_{ij}K^{ij}-K^2),
\end{eqnarray}
and consider
\begin{eqnarray}
S_{EHaW}=S_{EH}+S_{aWf},\label{nconformal}
\end{eqnarray}
where $S_{aWf}$ is the anisotropic Weyl action given in Eq.
(\ref{cconformal}) with an arbitrary potential $V(\varphi)$. Let us
introduce the Friedmann-Robertson-Walker metric via
\begin{eqnarray}
ds^2=-dt^2+a^2(t)\left[\frac{dr^2}{1-\bar{\kappa}
r^2}+r^2(d\theta^2+\sin ^{2}\theta d\phi^2)\right],
\end{eqnarray}
where $\bar{\kappa}=-1,0,1$ corresponding to an open, flat, closed
universe respectively. Note that for $z=1$, $\eta=1$, and $\xi=0$,
the action (\ref{nconformal}) is reduced to the conformal
quintessence~\cite{birrell,qu1,wetterich}. In this background, one
finds the following results,
\begin{eqnarray}
K_{ij}&=&-Hg_{ij}~,~~K=-3H,\\
R_{ij}&=&\frac{2\bar{\kappa}}{a^2}g_{ij}~,~~R=\frac{6\bar{\kappa}}{a^2}.
\end{eqnarray}
For the above action (\ref{nconformal}), the equations of motion
become
\begin{eqnarray}
\delta_{N}S_{EHaW};&&\frac{1}{2\kappa^2}(R-K_{ij}K^{ij}+K^2)+\delta_{N}S_{aWf},
\label{Neq1}\\
\delta_{N^i}S_{EHaW};&&\frac{1}{2\kappa^2}(-2\nabla_{j}K^{j}_{i}+2\nabla_{i}K)+
\delta_{N^i}S_{aWf},\label{Nieq1}\\
\delta_{g^{ij}}S_{EHaW};&&\frac{N}{2\kappa^2}A_{ij}^{(0)}+
\delta_{g^{ij}}S_{aWf},\label{gijeq1}
\end{eqnarray}
where
\begin{eqnarray}
A_{ij}^{(0)}&=&R_{ij}-\frac{1}{2}g_{ij}R
-\frac{\nabla_{i}\nabla_{j}N}{N}+\frac{\nabla_{k}\nabla^{k}N}{N}g_{ij}-
\frac{N_i\nabla_{k}K_{j}^{k}}{N}-\frac{N_j\nabla_{k}K_{i}^{k}}{N}+
\frac{\nabla_{i}N^{k}K_{jk}}{N}+\nn\\
&&\frac{\nabla_jN^{k}K_{ik}}{N}+\frac{N_{k}\nabla^{k}K_{ij}}{N}
-2K_{i}^{k}K_{jk}-\frac{1}{2}K_{kl}K^{kl}g_{ij}+
KK_{ij}-\frac{\dot{K_{ij}}}{N}\nonumber\\
&&-\frac{K^2}{2}g_{ij}+\frac{\nabla_{j}KN_{i}}{N}
+\frac{\nabla_{i}KN_{j}}{N}-
\frac{\nabla_{k}KN^{k}g_{ij}}{N}+\frac{\dot{K}g_{ij}}{N}.
\end{eqnarray}
Assuming a homogeneous scalar field $\varphi(x)=\varphi(t)$, one
obtains the following equations from Eqs. (\ref{Neq1}) $\sim$
(\ref{gijeq1}) and (\ref{phieq}):
\begin{eqnarray}
&&\hspace*{-5.0em}(\frac{1}{2\kappa^2}\varphi^{2z-6}+\eta)H^2=\eta
H\theta-\eta\frac{\theta^2}{4}+
\frac{1}{6}{\varphi}^{2z-6}V+\frac{\rho}{6}\varphi^{2z-6}
-\frac{\bar{\kappa}}{a^2}\left(\frac{\varphi^{2z-6}}{2\kappa^2}+\eta\varphi^{4z-4}
\right),\label{hhh1}\\
&&\hspace*{-5.1em}(\frac{1}{2\kappa^2}\varphi^{2z-6}+\eta)\dot{H}=-\frac{\varphi^{2z-6}}{4}
(\rho+P)+\eta\frac{\dot{\theta}}{2}+\frac{\eta z}{2}
(-H\theta+\frac{\theta^2}{2})+
\frac{\bar{\kappa}}{a^2}\left(\frac{\varphi^{2z-6}}{2\kappa^2}+\eta\varphi^{4z-4}
\right),\label{hhh2}
\end{eqnarray}
and
\begin{eqnarray}
\hspace*{-0.6em}\eta(z+1)\frac{\bar{\kappa}}{a^2}\varphi^{4z-4}+\eta(z+3)H^2-3\eta
H\theta
-\frac{\eta(z-3)}{4}\theta^2+2\eta\dot{H}-\eta\dot{\theta}-\frac{1}{12}
\varphi^{2z-5}V^\prime=0,\label{hhh3}
\end{eqnarray}
where $^\prime$ denotes differentiation with respect to $\varphi.$
Note that Eq. (\ref{Nieq1}) is satisfied trivially. When $\eta=V=0$,
one can check that Eqs. (\ref{hhh1}), (\ref{hhh2}) are equivalent to
the ordinary Friedmann equations. We assume flat universe with
$\bar\kappa=0$ from here on. Then, Eqs.
(\ref{hhh1})$\sim$(\ref{hhh3}) can combine into the following
equation,
\begin{eqnarray}
(z+3)H^2+2\dot{H}&=&\kappa^2\left(\frac{z}{3}\rho-P\right)-
\frac{\kappa^2}{3}\left(\frac{\varphi}{2}V^{\prime}(\varphi)-(z+3)V(\varphi)\right).\label{hhh7}
\end{eqnarray}
In the next subsections, we discuss possible solutions of these
equations for both with or without anisotropic Weyl invariance.

\subsection{Case with anisotropic
invariance } In the case of cosmological fluid with anistotropic
invariance, introducing $P=\bar{\omega}\rho$ then $z=3\bar{\omega}$.
Here $\bar{\omega}$ is the equation of state parameter. Since the
right hand side of Eq. (\ref{hhh7}) vanishes in this case, we have
\begin{eqnarray}
(z+3)H^2+2\dot{H}=0.\label{constr1}
\end{eqnarray}
In the above equation, one can check easily that in the case of $z\neq -3$, $H$
behaves as $1/t$ which correspond the power law solutions
and for $z=-3$, $H$ is a constant value which represents the exponentially
accelerating solution.

To see these behaviors in more detail,  we find explicit solutions of Eqs.
(\ref{hhh1})$\sim$(\ref{hhh3}). For power law solutions, we have
\begin{eqnarray}
\rho=\rho_0 a^{-3(1+\bar{\omega})},~~a=a_0
t^{\frac{2}{3(1+\bar{\omega})}},~~\varphi=\varphi_0
t^{-\frac{1}{3(1+\bar{\omega})}},~~\alpha=0,\label{zn3}
\end{eqnarray}
where $\rho_0,a_0,\varphi_0$ are positive constants which satisfy
the relations
$\rho_0a_{0}^{-3(1+\bar{\omega})}=\frac{4}{3\kappa^2(1+\bar{\omega})^2}$.
This solution preserves the standard cosmology. Most of all, in this
case, it should be remarked that $z=1,~0$ correspond to
$\bar{\omega}=1/3$ (radiation dominance),$~0$ (matter dominance)
respectively. In the region $-3<z<-1$, it has an accelerating phase,
whereas decelerating phase in $ z> -1 $. In particular, for $z=-3$
the solution is
\begin{eqnarray}
\rho=P=0,~~a=a_0 e^{Ht},~~\varphi=\varphi_0
e^{-\frac{H}{2}t},\label{z1}
\end{eqnarray}
where $H^2=\kappa^2\alpha/3$.

Note that with the anisotropic Weyl invariant fluid with
$P=z\rho/3$, the universe is described by different values of the
critical exponent in radiation-dominated, matter-dominated, and
vacuum-dominated epochs, but still can maintain the anisotropic Weyl
invariance\footnote{Note that the critical exponent  decreases as
the universe evolves. One possible interpretation of the result
might be that this running behavior of the parameter $z$ in the
Lagrangian  is due to the renormalization properties of the scalar
field theory and could be viewed as being reasonable in the sense of
the renormalization group.  However,  a demonstration of this
running behavior is beyond the scope of the present paper, and since
the macroscopic equations of state in each epoch is involved with
different cosmological fluids, it may be hard that it could actually
be realized  at the microscopic level. Another interpretation is
that there are several anisotropic conformal invariant sectors
described by $z$ and each epoch corresponds to one of these sectors
via cosmological fluid contents at that epoch.}.

However, if we break the anisotropic Weyl invariance of the
cosmological fluid, while preserving it for the potential part, we
can describe the extrapolation from matter dominance to vacuum
dominance of the universe with a single value of the critical
exponent.
 In this case, with $P=\bar{\omega}\rho$, Eq.(\ref{hhh7}) becomes
\begin{eqnarray}
(z+3)H^2+2\dot{H}=\frac{\kappa^2}{3}(z-3\bar{\omega})\rho.
\end{eqnarray}
In particular for $z=-3$,
\begin{eqnarray}
\dot{H}=-\frac{\kappa^2}{2}(1+\bar{\omega})\rho.
\end{eqnarray}
Then, from Eqs. (\ref{hhh1})$\sim$(\ref{hhh3}) we find the following
solution
\begin{eqnarray}
H=\sqrt{\frac{\kappa^2\alpha}{3}}\coth[At],~~\varphi=
B\left(\sinh[At]\right)^{-\frac{1}{3(1+\bar{\omega})}},~~
\rho=\alpha(\sinh[At])^{-2},
\end{eqnarray}
 where $A=(1+\bar{\omega})\sqrt{3\kappa^2\alpha}/2$, $B$ is a constant
 and $a(t)=a_0(\sinh[At])^{2/3(1+\bar{\omega})}$.
 For early universe with small t, $ a(t)\sim t^{2/3(1+\bar{\omega})}$, and it describes the standard
 power  law expansion with equation of state parameter with $\bar{\omega}$.
For late time with large $t$, $a(t)\sim
e^{\sqrt{\frac{\Lambda}{3}}t}$ with $\Lambda=\kappa^2\alpha$.  This
solution describes an extrapolation from matter-dominance into
vacuum dominance.

\subsection{Case without anisotropic invariance}
 In this case, $P=\bar{\omega}\rho$ and $V(\varphi)$ is different from the anisotropic invariant
 potential $\alpha\varphi^{2(z+3)}
$. Eq.(\ref{hhh7}) can be solved in three cases.

In the case $\rho=0$, there are two potential forms that are
interesting from the cosmological point of view.

 i)~$
V(\varphi)=A_1\varphi^{-2z+6}+A_2\varphi^{-2z+10}$,  where
$A_1,~A_2$ are some constants. Then
Eqs.(\ref{hhh1})$\sim$(\ref{hhh3}) and Eq.(\ref{hhh7}) yields the
following solution:
\begin{eqnarray}
H=B_1\tanh[B_2 t],~~\varphi=(-2\eta\kappa^2)^{\frac{1}{2z-6}},
\end{eqnarray}
 where $B_1=\sqrt{A_2/(3\eta(z+3))}(-2\eta\kappa^2)^{1/(z-3)},~B_2=\sqrt{A_2(z+3)
 /(12\eta)}(-2\eta\kappa^2)^{1/(z-3)}$ and
 $A_1=-(-2\eta\kappa^2)^{2/(z-3)}A_2$. Here $\eta<0$, $z~(\neq 3)>-3$ and $A_2<0$.
 It is interesting to note that in this case, the effective equation
 of state is given by
\begin{eqnarray}
\omega(t)=-1-\frac{z+3}{3}{\rm csch}^2[B_2 t]~<-1
\end{eqnarray}
corresponding to the phantom model \cite{copeland}.

 ii) ~$
V(\varphi)=\left(\frac{\varphi^{2z-6}}{1+\frac{1}{2\eta\kappa^2}\varphi^{2z-6}}
\right)^{\frac{z+3}{z-3}}.$
 With $\varphi=\varphi_0=const$, we find the de Sitter solution, $a\sim
e^{Ht}$, where
\begin{eqnarray}
H=\sqrt{\frac{1}{6\eta}}\left(\frac{\varphi_0^{2z-6}}{1+\frac{1}
{2\eta\kappa^2}\varphi_0^{2z-6}}\right)^{\frac{z}{z-3}}.
\end{eqnarray}
In the case of $z=3$ the above solution diverges and is replaced
with
\begin{eqnarray}
V(\varphi)=\varphi^{\frac{12}{1/2\eta\kappa^2+1}},~~H=\frac{1}{6\eta(1+1/2\eta\kappa^2)}
\varphi_0^{\frac{12}{1/2\eta\kappa^2+1}}.\nn
\end{eqnarray}

When $\rho\neq 0$, the available solution is given when
$V=V_0=const$ and $z\neq -3$:
\begin{eqnarray}
H=\sqrt{\frac{\kappa^2V_0}{3}}\coth[Ct],~~\varphi=D\left
(\sinh[Ct]\right)^{-\frac{1}{3(1+\bar{\omega})}},~~
\rho=V_0(\sinh[Ct])^{-2},
\end{eqnarray}
 where $C=(1+\bar{\omega})\sqrt{3\kappa^2 V_0}/2$, $D$ is a constant
 and $a(t)=a_0(\sinh[Ct])^{2/3(1+\bar{\omega})}$. This is also an
 extrapolating solution from matter dominance to vacuum dominance.

\section{Conclusion and discussion}
In this paper, we were able to construct an anisotropic Weyl
invariant action in the ADM formalism with the help of extra scalar
field, generalizing the original work of Ho\v{r}ava. Among possible
applications of the result, we treated this action as an anisotropic
Weyl matter field and considered the EHaW action adding the
cosmological fluid sector, and studied cosmological consequences. It
is found that if the anisotropic Weyl invariance is imposed, there
must be a correlation of the critical exponent $z$ and the equation
of state parameter $\bar{\omega}$ satisfying $\bar{\omega}=z/3$.
According to this condition, it is possible to reinterpret
cosmological evolution not by $\bar{\omega}$ but by $z$. In the
early universe, radiation and matter dominance correspond to $z=1$
and $z=0$ anisotropic Weyl invariance respectively. At late times,
it has $z=-3$ which is de-Sitter phase, i.e., $\bar{\omega}=-1$. It
is also shown that for particular value of $z=-3$, the potential
term for the scalar field becomes cosmological constant and there
exists an extrapolating solution from matter dominated epoch into a
late time accelerating universe in the broken cosmological fluid
case. We also found de Sitter solution in the case where the
anisotropic conformal invariance is broken by the potential term,
and especially in the polynomial potential (case i) of Sec. 4.2),
the effective equation of state parameter is less than -1. The
compatibility of the cosmology considered in this work with the
observed CMB anisotropies and structure formation remains to be
seen.

We recall that in the standard cosmology, the cosmological fluid
sector breaks conformal invariance, unless $\bar{\omega}=1/3$, i.e.,
radiation dominated. In the present anisotropic Weyl invariance,
cosmological fluid sector can maintain the invariance for an
arbitrary value of $\bar{\omega}$ due to the constraint
$\bar{\omega}=z/3$. However, for realistic cosmology, the
anisotropic conformal invariance has to be broken, and the parameter
$z$ is free. The physical significance of this parameter, in general
including the cosmological case is yet to be explored. One example
of anisotropic local conformal invariance appears in the condensed
matter system\cite{henkel}, and  considering the AdS/CMT
correspondence might shed some light on this.

We conclude with a couple of remarks on the issues related to
Ho\v{r}ava-Lifshitz gravity in the case of the anisotropic  Weyl
invariant action. The first one is the question on the possible
existence of the scalar graviton which shows pathological behavior.
In the original Ho\v{r}ava-Lifshitz gravity, this was pointed out to
cause serious problems, but, subsequently it was shown that this
could be cured via a natural extension of the Ho\v{r}ava-Lifshitz
gravity by abandoning the projectability and adding suitable space
dependent lapse functions \cite{blas}.   Since local Lorentz
invariance is broken when $z\neq 1$, the action (\ref{conformal})
confronts the same problem and the scalar graviton persists in the
anisotropic Weyl action (\ref{conformal}). To see this more closely,
we first fix the gauge by choosing a constant value for the field
$\varphi$. Then,  one can show that the $\eta$ term  in the action
(\ref{conformal}) does not produce any scalar graviton mode since
the ratio of $B_{ij}B^{ij}$ to $B^2$ is equal to $1$. Also, for the
$\xi$ term with Weyl invariance of the original Ho\v{r}ava-Lifshitz
gravity, it has been shown explicitly that this $\xi$ term also does
not produce graviton mode \cite{park}. Even though these terms do
not propagate the scalar graviton separately, their sum do propagate
the extra mode and the conformal action (\ref{conformal}) turns out
to coincide with the low-energy limit of the non-projectable
Ho\v{r}ava-Lifshitz gravity which was shown to propagate extra
graviton mode \cite{scalargraviton}. Unfortunately, this scalar
graviton problem is not cured for the full Einstein-Hilbert
anisotropic Weyl action (\ref{nconformal}). In this case, conformal
symmetry is not present and we cannot gauge fix $\varphi$ away.
Assuming that the action admits a flat background with a constant
$\varphi=\varphi_0$ which would require the potential to have a
minimum with $V(\varphi_0)=0$, we can consider a perturbation around
$\varphi_0$, $\varphi=\varphi_0+\tilde\varphi$. It turns out that
the Einstein-Hilbert action has the effect of changing the
coefficient of the scalar graviton mode, but the perturbed action
reduces to the theory where the $\tilde\varphi$ is coupled with the
low-energy limit of the non-projectable Ho\v{r}ava-Lifshitz gravity
previously mentioned. To investigate further the nature of this
scalar-tensor-type theory and especially to check whether the scalar
graviton problem can be cured along the line of Ref. [28] are left
as  open problems. The other is to check whether the anisotropic
Weyl action (\ref{conformal}) could be derived using the detailed
balance condition\footnote{In Ref.\cite{park} it was argued that the
anisotropic Weyl invariant action of Ho\v{r}ava-Lifshitz gravity
might be derived from the detailed balance condition.} of Ref.
\cite{Horava}. This condition may also restrict the diverse terms
which will appear when the anisotropic Weyl invariance is exploited
to construct the higher curvature terms not considered in this work.
This remains as a future study.

\section*{Appendix}
In the action (\ref{cconformal}), $S_{aWf}=S_{aW}+S_f$ is invariant
with respect to \bea N&\rightarrow&
e^{z\omega}N,~~~N_{i}~~\rightarrow ~~e^{2\omega}N_{i},\nn\\
g_{ij}&\rightarrow& e^{2\omega}g_{ij},~~\varphi~~~\rightarrow~
e^{-\frac{\omega}{2}}\varphi.\nn \eea For the infinitesimal
$\omega$, one can find the followings, \bea \delta N&=&z\omega
N,~~~~~\delta N^{i}~~=~~0,\nn\\ \delta g^{ij}&=&-2\omega
g^{ij},~~\delta\varphi~~~=~-\frac{\omega}{2}\varphi.\nn \eea And for
the above transformation, $\delta S_{aWf}$ is
\begin{eqnarray}
\delta S_{aWf}(N,g^{ij},N^{i},\varphi)~~=~~0&=&\frac{\delta
S_{aWf}}{\delta N}\delta N+\frac{\delta S_{aWf}}{\delta
g^{ij}}\delta g^{ij}+\frac{\delta S_{aWf}}{\delta
N^{i}}\delta N^{i}+\frac{\delta S_{aWf}}{\delta \varphi}\delta \varphi\nn\\
&=&\omega\left(zN\frac{\delta S_{aWf}}{\delta N}-2g^{ij}\frac{\delta
S_{aWf}}{\delta g^{ij}}-\frac{\varphi\delta S_{aWf}}{2\delta
\varphi}\right).\label{del}
\end{eqnarray}
Substituting (\ref{Neq}) $\sim$ (\ref{phieq}) into (\ref{del}) and
after some tedious calculations one can find the following
condition,
\begin{eqnarray}
&&2zN(-\rho)+6N P=0\nn\\
&\rightarrow& P=\frac{z}{3}\rho.\nn
\end{eqnarray}

\section*{Acknowledgments}
We  thank the anonymous referee for valuable suggestions. We also
like to thank Seyen Kouwn, Joohan Lee, Tae Hoon Lee, and Won-Il
Myeong for useful comments.  This work was supported by the National
Research Foundation of Korea(NRF) grant funded by the Korea
government(MEST) through the Center for Quantum Spacetime(CQUeST) of
Sogang University with grant number 2005-0049409.

\end{document}